\documentclass{elsart}
\newcommand{\xs}{ {\bf {x_s}}}

\newcommand{\x}{ {\bf {x}}}
\newcommand{\dd}{ {\bf {d}}}

\usepackage{epsfig}

\usepackage{amssymb}

\begin{document}

\begin{frontmatter}

\title{
Spatial
methods for
event reconstruction
in CLEAN
}

\author{Kevin J. Coakley\thanksref{label2}\corauthref{cor1}}
and
\ead{kevin.coakley@nist.gov}
\author{Daniel N. McKinsey\thanksref{label3}}
\ead{daniel.mckinsey@yale.edu}
\corauth[cor1]{
Contributions of NIST staff to this work are not subject to copyright laws in the US.
}

\address[label2]{Statistical Engineering Division, National Institute of Standards and Technology, 325 Broadway,
Boulder C0 80305
}
\address[label3]{Department of Physics, Yale University,
New Haven, CT 06520}

\author{}

\address{}

\begin{abstract}
In CLEAN
(Cryogenic Low Energy Astrophysics with Noble gases),
a proposed neutrino and dark matter detector,
background
discrimination
is possible 
if one can
determine the location of an ionizing radiation event with high accuracy.
Here, we develop spatial methods for event reconstruction,
and study their performance
in computer simulation experiments.
We simulate
ionizing radiation events that produce multiple scintillation photons 
within a spherical detection volume filled with
liquid neon.
We
estimate the radial location of a particular ionizing radiation event based
on the observed count data corresponding to that event. The
count data are collected by detectors
mounted at the spherical boundary of the detection volume.
We neglect absorption, 
but account for 
Rayleigh scattering.
To account for wavelength-shifting of the scintillation
light, we assume that photons are absorbed and re-emitted 
at the detectors. In our study, the detectors incompletely
cover the surface area of the sphere.
In the first method, 
we estimate the radial location
of the event
by maximizing the approximate Poisson likelihood
of the observed count data.
To correct for scattering and wavelength-shifting, we
adjust this estimate using a polynomial calibration model.
In the second method, we predict the radial location of the
event as a polynomial
function of the magnitude of the centroid of the observed count data.
The polynomial calibration models are
constructed from calibration (training) data.
In general, the Maximum Likelihood method estimate is more accurate than
that of 
the centroid method estimate.
We estimate the expected number of photons
emitted by the event
by a
Maximum Likelihood method and a simple method based on the
ratio of the
number of detected photons and a detection probability factor.
\\

\end{abstract}

\end{frontmatter}


\normalsize
\section{Introduction}

We estimate the location of an ionizing radiation event
that produces multiple photons within a spherical detection volume
based on count data recorded by detectors mounted on the
boundary of this detection volume.
The detectors cover approximately 75 $\%$ of the total
area of the detection volume boundary. 
We
assume that all detected photons produced by a particular
event can be distinguished according to their arrival times
from photon counts produced by other events.
Beyond this, we do not require additional
temporal information.
We assume that absorption is negligible, and that
photons undergo Rayleigh scattering.
We also estimate
the expected number of emitted photons 
for the event.

Our ``event reconstruction" work is motivated by
a proposed experiment called
Cryogenic Low Energy Astrophysics with Noble gases (CLEAN).
In CLEAN [1,2],
events would be detected based on
scintillation light produced by neutrino-electron
scattering and WIMP-nuclear elastic scattering
in a large cryostat filled with liquid neon.
In CLEAN, the expected number of scintillation photons
produced by an event would be proportional to the
amount of energy deposited by the event.
The proportionality factor would be determined from
a calibration experiment.
Such events of interest would occur uniformly throughout the
cryostat.
Here, we consider a cryostat with a spherical geometry.
Because neon has lower binding energy to surfaces
than most radioactive impurities, 
there should not be 
internal backgrounds in
the neon provided that the neon is purified using cold traps.
Background gamma ray events
would also produce scintillation light, and tend to occur near the
walls.
Near the center of a spherical detection
volume, the penetration probability 
of background gamma rays is very low.
Thus,
if 
the radial position of
an event 
can be determined accurately,
background gamma ray events can be discriminated
with high confidence
from events
of interest.

In this current paper, we focus solely on
the event reconstruction problem.
In a forthcoming paper, we will present
a detailed simulation of CLEAN, including
background gamma ray propagation.
In this later paper,
we will
quantify 
the performance 
of our event reconstruction method for determining
the location of an event in the context of the
physics goals of a CLEAN experiment.
We will also discuss
experimental design, calibration, 
and construction
of CLEAN.

Temporal methods, not spatial methods, are commonly used for event reconstruction
in the current neutrino experiments
KamLAND [3,4] and
Borexino [5]. 
Both of these experiments use organic scintillators.
Here, we 
present a spatial method 
for event reconstruction and carefully
study its performance in computer simulation experiments.
We remark that 
the XMASS [6]  
research team is also studying
spatial
event reconstruction methods for their proposed neutrino experiment.
Liquid xenon would be used as a scintillator in XMASS.
In the field of tomography,
spatial methods 
are commonly used to reconstruct
spatially varying intensity
images [7-11]. 
For each pixel or voxel, the corresponding intensity parameter
is the expected number of photons emitted during the 
experiment.
Instead of reconstructing a spatially varying
photon intensity image,
we seek to determine the intensity and location of a single source of
photons.
Thus, our problem is a special case of a more general
imaging problem.
In this work, we model the observed count data as 
a realization of a spatial Poisson process.
This approach is frequently taken to solve
a variety of imaging problems
in nuclear medicine [8-11].
We estimate the event location by
maximizing the
approximate
likelihood function of the observed detector count data
assuming a scatter-free transition matrix.
Because the transition matrix does not account
for scattering,
the estimate of the radial location of the
event can be strongly biased.
However,
we significantly reduce this bias, that is, systematic error,
by correcting the initial estimate based on
a
calibration model determined from calibration (training) data.
This is possible because the bias of the estimate introduced
by misspecification of the transition matrix has
a very predictable structure.

We do not mean to imply that bias results only because
of misspecification of the transition matrix.
For the case where we use the exact transition matrix, we expect some
bias because nonlinear estimation methods, including the
Maximum Likelihood (ML) method, generally produce biased estimates.
We expect this bias to be more significant
for low count situations
than for high count situations.
For a general discussion of this point, see [12].
For a specific illustration of this point, see [13].

We stress that the methods we develop here are appropriate
for the case where absorption of scintillation photons
within the neon is negligible. For the case where
absorption is significant, our methods
are not directly 
applicable without further modification.
If 
absorption effects are significant in the actual CLEAN
experiment, 
we plan to directly
estimate a transition matrix that will account
for scattering, absorption,
and other geometric effects.
(This direct transition matrix modeling approach has been
suggested in [6] for the planned XMASS experiment.)
Even if we have exact knowledge of the true
transition matrix, further calibration experiments
may be necessary
to quantify the 
performance of CLEAN.
The need for further calibration studies may be
most pressing for low count situations.

In Section 2, we  present the scattering model.
In Section 3, we 
present a Poisson likelihood model for
the data based on a scatter-free transition matrix.
In Section 4, we study the statistical properties
of both the uncorrected and corrected radial estimates
for a variety of cases.
For the same data, we compare a prediction model based on 
the centroid of the observed counts with the ML estimate.
We also estimate the expected number of 
emitted photons at the event location using the ML method
and a simpler method. In the simpler method, 
the estimate is proportional to the number of detected photons.

\section{Simulation Model}

At a particular location, multiple scintillation photons
are produced by an event of interest.
The probability density function
(pdf) for the initial velocity direction of each emitted photon is
uniform on the surface of the unit sphere.
Following standard practice [14,15] 
we simulate the distance traveled before first scattering
(as well as the distance between
subsequent scatterings)
by sampling from an exponential distribution.
The
expected value of this
realization
is the scattering length $\lambda_s$.
At the point corresponding to the
first scattering, we rotate the velocity direction  
about its original direction according to a model
for the differential cross section of the Rayleigh scattering
process.
We neglect angular variation of the atomic form factor. Thus,
the inner product
of the new velocity direction and the old velocity direction,
$\cos( \theta)$,
has a pdf
proportional to
$(~1 ~+~ \cos^2(\theta)~)$.
We select a position on the cone defined by
$\cos( \theta)$ and the original velocity direction by
sampling an azimuthal angle $\phi$ that is uniformly
distributed between 0 and $2 \pi$.

We compute the point where a photon crosses the
spherical boundary of the detection volume.
If this point is not within the area
of a detector, the photon is not detected.
For the case where
the crossing point is within the
area covered by a detector element,
we consider two detection models.
In the ``no-shift" model, the photon is detected by the detector element
with probability $p_e$. We assume that
$p_e$ is
independent of the photon trajectory with respect
to the detector. 
In the ``shift" model, the detector element absorbs the photon and 
converts it to lower energy. This wavelength-shifted photon is randomly re-emitted.
We study the ``wavelength-shifting" detection model because the extreme
ultraviolet scintillation light 
that would be
produced by liquid neon
in CLEAN is not directly detectable by conventional detectors.
However, 
after being shifted to a longer wavelength,
this light is detectable by conventional detectors.
The corresponding pdf of the velocity direction of the shifted photon is uniform
on the surface of the unit sphere.
If the shifted photon's velocity points radially outward, the 
re-emitted photon is detected with probability $p_e$ at the detector element where shifting occurs.
Otherwise, the
photon 
travels inward, without scattering, until it crosses the spherical
detection boundary.
If this crossing  point 
falls within the area covered by
a detector element, it is detected
with probability $p_e$.
Otherwise, the photon is absorbed by the spherical wall and lost.
Our 
assumption that the wavelength 
shifted photon
does not scatter
is based on the fact that the shifted photon 
has much longer wavelength (about 400 nm)
than the original scintillation photon (about 80 nm),
and Rayleigh scattering is much stronger at shorter
wavelengths.

\section{Estimation Method}

\subsection{
Likelihood Model}

Given that a photon is emitted at location $\xs$,
the
probability that the $k$th detector
will detect this photon
is
denoted by
$P(k|\xs,p_e)$ where $p_e$ is the efficiency of the detector.
By varying $k$ and $\xs$, we call the set
${
P(k|\xs,p_e)
}$
the probability transition matrix.
For the ``no-shift"
detection model where 
there is no scattering, that is,
$\lambda_{s} ~=~ \infty$,
the transition matrix is
\begin{eqnarray}
P(k|\xs,p_e) =\frac{p_e}{4 \pi } \int_{A_k}^{}  \frac{ (\x - \xs) \cdot \x }
{ | \x - \xs  | ^ 3  | \x | }  d \x,
\end{eqnarray}
where we integrate over the area of the $k$th detector.
The expected number of detected photons at the $k$th detector 
is
\begin{eqnarray}
< n_k> ~=~\lambda_{k}(\xs) = \lambda P(k|\xs,p_e),
\end{eqnarray}
where the intensity
parameter 
$ \lambda $ 
is the expected number of photons emitted during the experiment.
We approximate the integral in Eq. 1 as
\begin{eqnarray}
P(k|\xs,p_e)
~\approx~ \frac{p_e p_{c} R^2}{N_{det}
| \dd_k - \xs  | ^ 2
} \cos ( \dd_k - \xs, \dd_k),
\end{eqnarray}
where $\dd_k$ is the point where the $k$th detector
is tangent to the sphere,
$p_{c}$
is the
fractional area of the surface of the sphere covered by
all detectors, and
$N_{det}$
is the number of detectors.
We simulate data for 
$N_{det}=$
2072  detectors tangent to the
spherical boundary of the detection volume
(Figure 1). We determine the tangent points
by an optimal packing scheme [16].

The actual number of emitted
photons at the event location is a realization of a Poisson process
with intensity parameter $\lambda$.
The expected value of the realization equals the intensity parameter.
We assume that
the number of counts at the $k$th detector
is a realization of a Poisson process with intensity parameter $\lambda_k$.
Thus,
the Poisson log-likelihood function of the observed data 
is
\begin{eqnarray}
\log L =  \sum _ { k} ^{N_{det}}
-\lambda_{k}(\xs) ~+~ n_{k} \log(\lambda_{k}(\xs)) -  \log( n_{k} ! ),
\end{eqnarray}
where $n_k$ is the number of detections at the $k$th detector.
We estimate the model parameter vector
${\bf \theta} = (x,y,z,\lambda)$
by
maximizing Eq. 4.
Since, we use an inexact transition matrix, 
strictly speaking, we obtain an approximate ML
estimate of the model parameters.

For the case where the transition matrix is exact,
the $(i,j)$th component of
the asymptotic covariance 
matrix of the  ML model parameter is
\begin{eqnarray}
\widehat{COV}( \hat{\theta}_i,
\hat{\theta}_j)
~=~
( I ^ {-1} ) _ {ij},
\end{eqnarray}
where
the Information matrix
is
\begin{eqnarray}
 I   _ {ij}
=
\sum_{k=1}^{N_{det}}
\frac{1}{ \lambda_k }
\frac { \partial { \lambda_k(\xs) } }
{ \partial { \theta _ i } }
\frac { \partial { \lambda_k(\xs) } }
{ \partial { \theta _ j } }
.
\end{eqnarray}
\\
For the Poisson model,
${ VAR(n_k) }=  \lambda_k(\xs)$.
The asymptotic variance of the $k$th  parameter estimate 
is the $k$th diagonal element of the asymptotic covariance matrix.
The asymptotic standard error (s.e.) of the $k$ parameter estimate
is the square root of this asymptotic variance.

Our estimate of the radial location of the event is
\begin{eqnarray}
\hat{r} = \sqrt{ \hat{x}^2 + 
\hat{y}^2 + 
\hat{z}^2}
.
\end{eqnarray}
For the special case where
$\xs ~=~ (0,0,z)$ 
and there is no scattering,
we approximate the 
standard error of $\hat{r}$ as
\begin{eqnarray}
\sigma_{\hat{r}} \approx \sigma_{\hat{z}} ~=~ 
( I ^ {-1} ) _ {33}.
\end{eqnarray}

For more details about asymptotic
statistical theory,
and
a discussion of a closely related 
ML estimation problem 
involving Poisson data,
see [17] and [13].

\subsubsection{Efficient Simulation}

If the detector efficiency
$p_e$
is much less than 1,
many simulated
photons are not detected even if they 
hit a detector element. 
Below, we describe an efficient simulation
model which 
simulates the relevant fraction 
of 
emitted photons 
that are 
detected (provided that they
hit a detector in the ``no-shift" detection model
or in the ``shift-model" after wavelength shifting).
On average, the relevant fraction that we simulate
is $p_e$. 

In our simulation model,
we assume that
the efficiency of the detector
$p_e$
is independent of the photon trajectory with respect
to the detector.
Thus,
$ \lambda P(k|\xs,p_e)  = 
\lambda p_e P(k|\xs,1) $ (Eq. 2).
Because of this relationship, we need not
simulate trajectories for all simulated photons that are
emitted at $\xs$.
In our efficient simulation model, the number of
photons 
simulated is a realization of a Poisson process with
expected value $\lambda p_e$ rather than $\lambda$.
However, in the efficient simulation model, the detector
efficiency is 1 rather than $p_e$.
That is, for data simulated with the efficient approach,
we
replace
$P(k|\xs,p_e)$ with
$P(k|\xs,1)$.
In both simulation models,
the simulated number of detected
photons at the $k$th detector
is a realization of a Poisson process with
intensity parameter
$\lambda_k$.
Because of this substitution,
we estimate a modified intensity $\lambda p_e$ rather than
the acutal intensity $\lambda$.
Throughout this work, we use this efficient approach.

\subsubsection{Optimization Details}

We seek the values of
the model parameters that yield the global maximum
(assuming that it is unique)
of
the approximate log-likelihood function, $\log L$ (Eq. 4), using an iterative algorithm.
Since optimization codes search for the global minimum
of a cost function, we actually minimize $-\log L $.
In each iteration, there are 3 steps. 

1. Select initial values of parameter estimates.

2. Estimate model parameters by minimizing $-\log L$
subject to the constraint that
the event location is within the
spherical detection volume.

3. Search for the global
minimum of $-\log L$
using the estimates from step 2 as initial values.

In step 1,
for the first iteration,
the initial estimate of  the
location of the event is the centroid of the observed count data.
The initial estimate of the modified intensity parameter $\lambda p_e$ is
the number of observed counts divided by the coverage 
factor ($p_c$).
For subsequent iterations,
the initial location
is
the sum of the
centroid 
and a random perturbation vector. Each component is
a normal (Gaussian) random variable with expected value equal to 0 and
standard deviation equal to 0.05 R.
We require that 
the
location vector has magnitude less than or equal to 0.95 R.
We perturb
the
initial estimate of $\lambda p_e$  by multiplying $n/p_c$
by a factor $1 + \epsilon$,
where
$\epsilon$ is a normal random variable with expected value 0 and
standard deviation 0.02.

In step 2,
we  minimize  the sum of
$-\log L $ 
and a penalty function,
using an algorithm based on
a quasi-Newton method [18].
The penalty function
forces the estimate to fall within
the detection volume.
The penalty function is 0 when $\hat{r} < R $, and
10 000 $(\hat{r}/R)^2$ when $\hat{r} \ge R $.

In step 3,
we minimize the cost function, $- \log L $,
using 
a modified Newton method where
we supply both the first and second derivatives
of the cost function
with respect to the model parameters
[19].

In general, the numerical stability of the estimate 
depends on the number of counts, the scattering length,
and the geometry of the detectors.
For low count situations, stability is usually achieved by
30 iterations. 
For instance,
for the case of wavelength-shifting and $\lambda_s/R = $ 0.1,
we simulate 500 events of 
interest for each of three cases. In the three cases, the number of
observed counts $n$ is 10, 25 and 50.
For $n =$ 10, the difference between the uncorrected estimate of $\hat{r}/R$
at 30 and 90 iterations
was greater than 0.01 only three times.
For $n=$25 and $n=$50, the 
difference between the estimate at 30 and 90 iterations
was less than 0.01 for all 500
realizations.
The number of cases 
for which the difference between the estimate at the  first and 90$th$ iterations
was greater
than 0.01 is
respectively
14, 1, and 4 for
$n=$ 10, 
$n=$ 25, 
and
$n=$ 50.
For all cases where the estimate
at the first and 90$th$ iterations varied by more than 0.01,
the ``instability" occurs at 
$\hat{r}/R >  $ 0.65.
Hence,
the outcome of a classification rule to determine if an
event is either ``background" or an ``event of interest"
is less sensitive to the number of iterations than is
the point estimate of radial location.
For all cases studied here,
stable estimates are obtained by 30 iterations, provided that
the number of detected photons is greater than 50.

\subsection{Calibration Model}
Our calibration model is
\begin{eqnarray}
\frac{\widehat{r_c}}{R} = 
\alpha_1 
 \frac{\hat{r}}{R} 
~+~ \alpha_2
 (\frac{\hat{r}}{R} )^2
~+~ \alpha_3
 (\frac{\hat{r}}{R} )^3,
\end{eqnarray}
where $\widehat{r_c}$
is the corrected estimate. We estimate the
calibration model parameters, $\alpha_1,\alpha_2,\alpha_3$,
from calibration data
by minimizing
\begin{eqnarray}
\sum_{i} ^ {}
( r_i ~-~
\alpha_1 
 \frac{\hat{r}_i}{R} 
~+~ \alpha_2
 (\frac{\hat{r}_i}{R} )^2
~+~ \alpha_3
 (\frac{\hat{r}_i}{R} )^3
)^2
,
\end{eqnarray}
where $r_i$ and $\hat{r}_i$
are the true and estimated values at the $i$th 
point in the calibration data set.
The radial locations of the points in the 
calibration data are
uniformly distributed
(Table 1).
We  quantify the accuracy of the calibration model by 
computing
RMSE = $\sqrt{\frac{m}{m-k}\sum_{i=1}^{m} (\frac{ \hat{r}_i}{R} -
- \frac{r_i}{R} ) ^ 2}$, where $m$ is the total number of points in the calibration
data set and $k$ is the number of calibration model parameters.
For the cases studied, $m \approx 2500 $.
In Table 1 and elsewhere, we represent the standard error, that is, the
estimated standard deviation,
of an estimate in parentheses. For instance, 0.894(16)
means  the estimated value is 0.894, and the associated standard
error is 0.016.

In this work, we focus on estimation of the radial location of an
event. If one wishes to estimate the Cartesian coordinates
of the event, we suggest that 
the estimates of $x$,$y$ and $z$ are corrected in the same way as
$\hat{r}$.  
That is,
${ \hat{x}_c } =
\frac{ \hat{r}_c } { \hat{r}  }  
 { \hat{x}  }, ~ 
{ \hat{y}_c } =
\frac{ \hat{r}_c } { \hat{r}  }  
 { \hat{y}  }, ~ 
$ and $
{ \hat{z}_c } =
\frac{ \hat{r}_c } { \hat{r}  }  
 { \hat{z}  }$.
In Figure 2, we illustrate how this approach works for
an example.

\section{Simulation Results}

\subsection{Position Estimation}

In our first study, we simulate data
at selected points on the
z-axis of our detector coordinate system for the ``no-shift"
detection model. For each point, we simulate
100 data sets.
The modified intensity parameter is 
$\lambda p_e = $ 200. 
For the ``no-shift" case,
the bias of the uncorrected 
estimate increases
as the scattering 
length decreases (Figure 3).
This bias is expected because, in general,
scattering 
increases the probability that a photon emitted in 
the upper hemisphere $z > 0$ will 
cross the spherical boundary in the upper hemisphere.
For instance, for the case where photons
are emitted at $\x=(0,0,0.9R)$,
the probability that the
crossing occurs at $z > 0.9 R$ is
0.541(5), 0.708(5), 0.801(4) for 
$\lambda_s/R = 1,0.1,0.01$.
This phenomenon
appears to be related 
to the behavior of
particles undergoing Brownian motion
near absorbing walls [20].

The bias of the corrected estimate
is significantly lower than the bias of
the uncorrected estimate. 
In general, the  
RMSE
of the corrected estimate 
decreases as the
scattering length decreases.
Moreover, the standard error of the corrected estimate
is often below the asymptotic standard error computed from
asymptotic theory for the scatter-free case.
This is not inconsistent with statistical theory.
Moreover, our estimation method
exploits additional information provided in the calibration
data that is not accounted for in the
asymptotic theory calculation.
We stress that our results correspond to the
case where absorption is negligible.
For the more general case where absorption effects are
significant, the accuracy of a corresponding prediction model may
decrease as the scattering length decreases.

For the case where there is no scattering, that is, $\lambda_s = \infty$, 
``shifting" introduces a negative bias
(Figure 4).
For the ``shift" model,
as the scattering length decreases, the bias increases in general.

In CLEAN, events of interest would occur uniformly
throughout the detection volume. Hence, the
pdf for the radial location
of an event of interest is proportional to $r^2$.
We simulate events of interest that produce a fixed number of counts $n$.
In Figure 5
we display scatterplots of the true value of $r$ and the
uncorrected estimated value of $r$ for
three cases.
In CASE A, we consider the ``shift"
detection model where $\lambda_s/R = \infty $.
In CASE B, we consider the ``no-shift"
detection model where $\lambda_s/R = 0.1 $.
In CASE C, we consider the ``shift"
detection model where $\lambda_s /R = 0.1 $.
In Figure 6,
we display scatterplots of the true value of $r$ and the
corrected estimated value of $r$ for
the same three cases.
In each scatterplot,
we show the 0.1 quantile of the
empirical distribution of the estimate as a dashed line
and the level $r_p/R = p ^ {1/3}$ where $p$ = 0.1 as a solid line.
If our estimate $\hat{r}$
were uniformly distributed in the spherical detection region,
the fraction of estimates less than
$r_p =  R p ^ {1/3}$
would be $p$.
Thus,
ideally, for each value of $n$,
one tenth of the estimates should fall below
$r_{0.1}$.

Assuming that events that produce a fixed number of observed photons
occur uniformly throughout the detection volume,
the expected value of the fraction of corrected estimates of $r$ below 
$r_{p}$ divided by $p$ is the efficiency of the detector.
Ideally, the detector efficiency should be 1 for all $n$.
However, 
our calibration
model (Eq. 9)
does not guarantee a detector efficiency 
equal to 1.
Since
a 
nonuniform detector
efficiency as a function of $n$
will distort the
count spectrum of the detected events, one should
quantify the detector efficiency.
As an illustration, we
estimate detector efficiency for
the ``shift" detection model where
$\lambda_s  / R = 2/9$ (Figure 7).

For experiments like CLEAN, we wish to determine whether an event
occurs within a particular inner spherical
region or fiducial volume defined by
$r <  R p ^ {1/3}$, where $0 < p < 1 $.
In principle, given enough calibration data,
we can determine the radial boundary of
a 100 $\times~ p$ percent fiducial volume so that 
the detector
efficiency is arbitrarily close to 1
by 
setting
$r_p$ equal to the $p$th quantile of
the distribution of estimated values of $r$.
This classification problem can be viewed as
a statistical hypothesis test where the null hypothesis is
that the actual radial location of a particular event 
is a realization from a distribution that has
a pdf proportional to $r^2$.
In this way of looking at the problem,
the estimated value of $r$ (corrected or uncorrected)
can be interpreted as the test statistic.
We accept the null hypothesis if the test statistic
falls below a certain critical level. 
The $p$th quantile of the distribution of this test statistic
(simulated under the assumption that the null hypothesis is
true) is the critical level for a test with size $1-p$
[21].
For a test
with size $1-p$,
when the null hypothesis is true,
we reject it with probability $1-p$.
When one
sets
the critical level of a test with size $1-p$
to the
$p$th quantile of a test statistic,
one has
calibrated a confidence point 
[22 page 263].
We wish to reject the null hypothesis with high probability
when an event is due to background.
That is, we wish the power of the test to be high.
When events are due to background,
a high rejection rate 
will yield a low background.

\subsubsection{ Centroid Method}
For comparison,
we predict the radial location of an event based on
the magnitude of the centroid of the count data.
The centroid $\vec{C}$ is defined to be
\begin{eqnarray}
\vec{C}~=~
\sum_{k}^{} w_k \dd_k,
\end{eqnarray}
where
\begin{eqnarray}
w_k ~=~
\frac{n_k}{\sum _{j=1}^{N_{det}} n_j }
\end{eqnarray}
and $n_k$ is the number of counts at the $k$th detector.
Our Centroid method estimate of the radial location of the
event is
\begin{eqnarray}
\frac{\hat{r}_{centroid}}{R} ~=~
\beta_1 \frac{ | \vec{C} | } { R} 
~+~
\beta_2 ( \frac{ | \vec{C} | } { R}  )  ^ 2
.
\end{eqnarray}
In general,
the variability of the Centroid method
estimate (Figure 8, Table 2)
is higher than that of the
ML method (Figure 6, Table 1).
The difference in performance is most dramatic for low count ($n < 50$) 
cases.

\subsection{Intensity Estimation}

In the efficient simulation model,
we simulate a Poisson random variable $N$
with expected value
equal to modified intensity
$\lambda p_e$.
To derive a simple estimate of the 
modified intensity, we 
assume that
the number of detected counts $n$,
given N,
is a binomial random variable with expected value $Np$ and
variance $Np(1-p)$.
For the ``no-shift" detection model 
$p=p_c$. 
For the ``shift" detection model,
$p={p_c ( 1 - 0.5 ( 1 - p_c) ) }$
.
Thus,
our ``naive" estimate  of the modified intensity is 
\begin{eqnarray}
\widehat{\lambda p_e}_{~naive} ~=~
\frac{n}{p}
.
\end{eqnarray}

Our second estimate of the modified intensity
is
\begin{eqnarray}
\widehat{\lambda p_e}_{ML} ~=~
\frac{p_c} {p}
\hat{\theta}_4
,
\end{eqnarray}
where 
$\hat{\theta}_4$ is the ML method estimate of $\lambda p_e$
determined by maximizing the log-likelihood of
the observed data (Eq. 4).
We adjust
$\hat{\theta}_4$
because our transition matrix
does not account for 
the effects of
wavelength-shifting
by the detectors.
For the cases studied,
for $r/R < $ 0.9,
the two methods yield similar
results. However, for $r\approx$ R, the
variability of
ML method estimate 
is generally higher than that of the
naive method estimate.
To illustrate this point, we compare the two methods
for the ``shift" detector
model where $\lambda_s/R $ = 0.1 (Figure 9). 

In an additional study,
we simulate
realizations of
$N$ that are uniformly distributed between 50 and 1200.
The Cartesian coordinates of each event location
are randomly generated so that
the
radial 
location of a random event is
uniformly distributed between 0 and $R$.
We compute the test statistic
\begin{eqnarray}
z ~=~ \frac{   n -  N p  } {\sqrt{ N p ( 1 - p) }}
\end{eqnarray}
for each event.
Assuming that our
binomial assumption is correct, the pdf of $z$ 
should be
well approximated by a unit normal (Gaussian) distribution centered on 0.
That is, the expected value and variance of $z$ 
should be
approximately
0 and 1.
Clearly,
for events near the wall,
the observed variability of $z$ is
inconsistent with the
binomial assumption
(Figure 10). We attribute this high variability
to the incomplete coverage of the walls by the detectors.
For instance,
in some cases
events can occur near the wall
in gaps between
detectors. For such cases, almost all of the emitted photons
traveling radially outward could be lost to the walls.
Assuming that 
the other half of the photons that travel radially inward
are detected with probability $p$,
the overall detection probability would be approximately
$p/2$ rather than $p$.
We attribute the apparent outliers for CASE A in Figures
5,6, and 7 to a similar wall effect.

For the ``no-shift" detection model,
the probability of detecting a photon that
is emitted at the center of the spherical detection volume
is $p_c = $ 0.75295(13).
Based on $p_c$, we predict the 
probability of detecting a photon emitted from the center of the spherical
detection volume to be $p=$ 0.65994(16) using 
the formula $p=p_c(1 - 0.5(1-p_c) )$ presented in Section 4.2.
For simulated events where the initial radial position
of the photon was uniformly distributed between 0 and 0.9 R,
we compute the ratio of detected to emitted photons for
for scattering lengths $\lambda_s/R~=~ 0.01, 0.1, 1, \infty$.
For the ``no-shift" detection model, this ratio is respectively
0.7536(3), 0.7526(3), 0.7529(3), and 0.7528(3).
For the ``shift" detection model,
this ratio is respectively
0.6607(3), 0.6605(3), 0.6606(3) and 0.6604(3).

\section{Discussion}

Since the variability of the Centroid method estimate of radial
location is higher
than the variability of the ML method estimate, we expect 
the ML method
to be 
a more powerful method for background discrimination
applications.
The relative performance
of
the two methods
should be most dramatic for low count situations.
Quantifying this relative performance 
is a subject for further study.

For an actual energy spectrum of interest,
due to wall effects, 
the actual distribution of events of interest that produce
$n$ observed events might deviate slightly from a uniform
distribution even if the actual distribution of
events that yield $N$
has a uniform distribution.
If this deviation from uniformity is significant, we should
account for it when constructing critical levels of
statistical tests to determine whether an event is due to a background
event.

It is reasonable to 
assume that the number of emitted scintillation photons
is a realization of a Poisson process with a rate parameter proportional
to the energy deposited by the event.
Imagine that some physical process of interest produces
events with energy deposit spectrum $f(E)$.
The spectrum $f(E)$ is a pdf in energy space.
Due to the Poisson assumption,
the number spectrum $g(N)$ generated by this process will
be proportional to
the
convolution of $f(E)$ with a transition matrix $p(N|E)$.
The spectrum for observed counts $q(n)$ is
\begin{eqnarray}
q(n) ~\propto~  \sum_{N}^{}
\int_{E}^{}
\int_{\xs}^{}
f(E)p(N|E)h(n|N,\xs)d \xs dE
,
\end{eqnarray}
where
$h(n|N,\xs)$
is a transition matrix that accounts for incomplete detector 
coverage and position effects like the wall effects
noted in this work.
Estimation of the true energy spectrum of interest $f(E)$
from the empirical count distribution
observed in an experiment is a challenging statistical inverse problem
worthy of further study.
However, it is possible to develop statistical tests to determine 
whether
a theory explains the observed
data without ``solving" the inverse
problem if,
given a theoretical estimate of $f(E)$,
we can
predict the count spectrum for the observed data.

\section{Summary}

We 
estimated the radial location of an ionizing radiation event
that produced multiple photons at a point
within a spherical detection volume,
based on the observed count data collected by detectors
mounted at the spherical boundary of the detection volume.
We neglected absorption 
but accounted for 
Rayleigh scattering as well as photon conversion
and re-emission at the detectors.
In one method, the predicted value was a polynomial
function of the centroid of the observed count data.
In the second method, 
the predicted value was a polynomial
function of the approximate maximum 
likelihood estimate of the radial location.
In both cases, the polynomial correction models were
constructed from calibration data.
In general, the ML method estimate was more accurate than
the Centroid method estimate.
In this work, our ML estimate
of location was determined
using a transition matrix
that neglects scattering and detector conversion of
photons.
We corrected our estimate of radial location 
in a post-processing step using a calibration model.
For the case of negligible absorption,
as the scattering length decreased,
the accuracy of the corrected estimate improved.
For the case where absorption is significant,
we may not observe this sort of improvement.

The event reconstruction methods 
presented in this work
do not account for
photon absorption in neon.
For the case where absorption is significant,
we would either modify our methods or
adopt a different method.
For the case where absorption is significant,
a spatial ML method based using estimate of
the true transition matrix, as suggested in [6],
is a promising approach. 
However,
as we mentioned earlier,
even if one has a
perfect estimate of the transition matrix 
accounting for scattering and other effects such
as absorption, the ML method estimate of
radial position may still be biased. 
Additional calibration experiments may be necessary
to quantify detector efficiency 
even if we know
the true transition matrix.

We estimated the
intensity of the event
using the ML method and a much 
simpler scheme.
In the simpler scheme, the estimate is
the number of detected photons
divided by a detection probability factor.
In general,
the simpler method was as good as or better than the ML method.
For the events near the wall, the variability of the ML estimate
was higher than the variability of the simple method.

In future work, we will study how well our event reconstruction
method discriminates events of interest from background events
for simulated data.
In particular, we plan to study the relative performance of the
Centroid and Maximum Likelihood methods.
We expect the relative performance of the two methods to be
most dramatic for low count situations.
For high count situations, 
both methods could be useful. For instance, the Centroid method
might serve to validate results obtained by the Maximum Likelihood
method.
In an actual experiment, multiple-point background events
can occur because background gamma rays
can deposit fractional amounts of their total energy.
Additional
simulation studies indicate
that
we can
discriminate
such multiple-point background events from single-point events of interest.

Our work demonstrates the feasibility of spatial
methods for event reconstruction
for the case where scattering is significant but absorption is negligible.
We developed calibration models to adjust estimates of
the radial 
location of an event based on simulated calibration data.
In a simulation, one can sample calibration data throughout the entire
detection volume. In a real experiment, one might not have such
freedom.
If sampled points are
not sufficiently representative of all points in the
detection volume,
the accuracy of the calibration model could be affected.
Thus, 
validation of empirical calibration methods 
for an actual experiment is an important
topic for further study.
Other important topics for further study include:
energy spectrum estimation,
quantification of detector efficiency and false detection
rate as a function of the number of detected counts,
and
development of
statistical hypothesis tests
for neutrino models.
\\

{\bf Acknowledgements}
We thank Adele Peskin of NIST for 
visualization  
work. We thank
Andrew Hime of Los Alamos National Laboratory and
Grace L. Yang of the University of Maryland
and NIST for useful comments.
The work of Daniel N. McKinsey
was supported in part by the National Science Foundation under
grant number PHY-0226142.
\vskip .12in

\newpage{}

\newpage{}
\begin {center}
\noindent {Table  1.
Calibration model parameters for Maximum Likelihood Estimate.
RMSE is the square root of the mean square
prediction error, adjusted for degrees of freedom, computed from all calibration data.
The number of counts in each calibration data set varies from 
50 to 1200. 
}
\\
\begin{tabular}{ccccc} \hline
$\lambda_s/R$ & $\widehat{\alpha_1}$ & $\widehat{\alpha_2}$ &$\widehat{ \alpha_3} $ & RMSE \\
\\ \hline
no-shift
\\
$\infty$ & 0.894(16) & 0.261(46)&  -0.156(33) &  0.0357
\\
1 & 0.779(15) & 0.312(41)& -0.072(28) & 0.0308
\\
0.1 & 0.602(14)& 0.264(39)& 0.182(26) & 0.0271
\\
0.01 & 0.618(9)& 0.047(25)& 0.368(17) & 0.0246
\\

shift
\\
$\infty$ & 1.316(20) &  0.023(55)&  -0.366(37) &  0.0501
\\
1 & 1.096(14) & 0.347(40) & -0.455(28) & 0.0435
\\
0.1 & 0.856(11) & 0.419(37) & -0.258(28) & 0.0344
\\
0.01 & 0.855(12) & 0.189(38) & -0.033(28) & 0.0328
\\
\hline
\end {tabular}
\end{center}

\newpage{}
\begin {center}
\noindent {Table   2.
Calibration model parameters for Centroid Estimate.
RMSE is the square root of the mean square
prediction error, adjusted for degrees of freedom, computed from all calibration data.
The number of counts in each training data set varies from 
50 to 1200. 
}
\\
\begin{tabular}{cccc} \hline
$\lambda_s/R$ & $\widehat{\beta_1}$ & $\widehat{\beta_2}$ & RMSE \\
\\ \hline
no-shift
\\
$\infty$ & 1.483(10) & 0.0108(194) & 0.0465
\\
1 & 1.326(6) & -0.0525(115) & 0.0370
\\
0.1 & 1.049(5) & 0.0167(64) & 0.0291
\\
0.01 & 0.991(4) & 0.0176(53) & 0.0257
\\
shift
\\
$\infty$ & 2.094(21) &  -0.113(56) &  0.0675
\\
1 &1.826(15) & -0.084(35)  & 0.0563
\\
0.1 & 1.475(8)  & 0.006(14)   & 0.0438
\\
0.01 & 1.386(7)  &  0.018(14)  &  0.0415
\\
\\ \hline
\end {tabular}
\end{center}

\newpage{}

\newpage{}
\begin{figure}
\vspace{0.01in}
\centerline{\epsfysize=4.0in
\epsffile{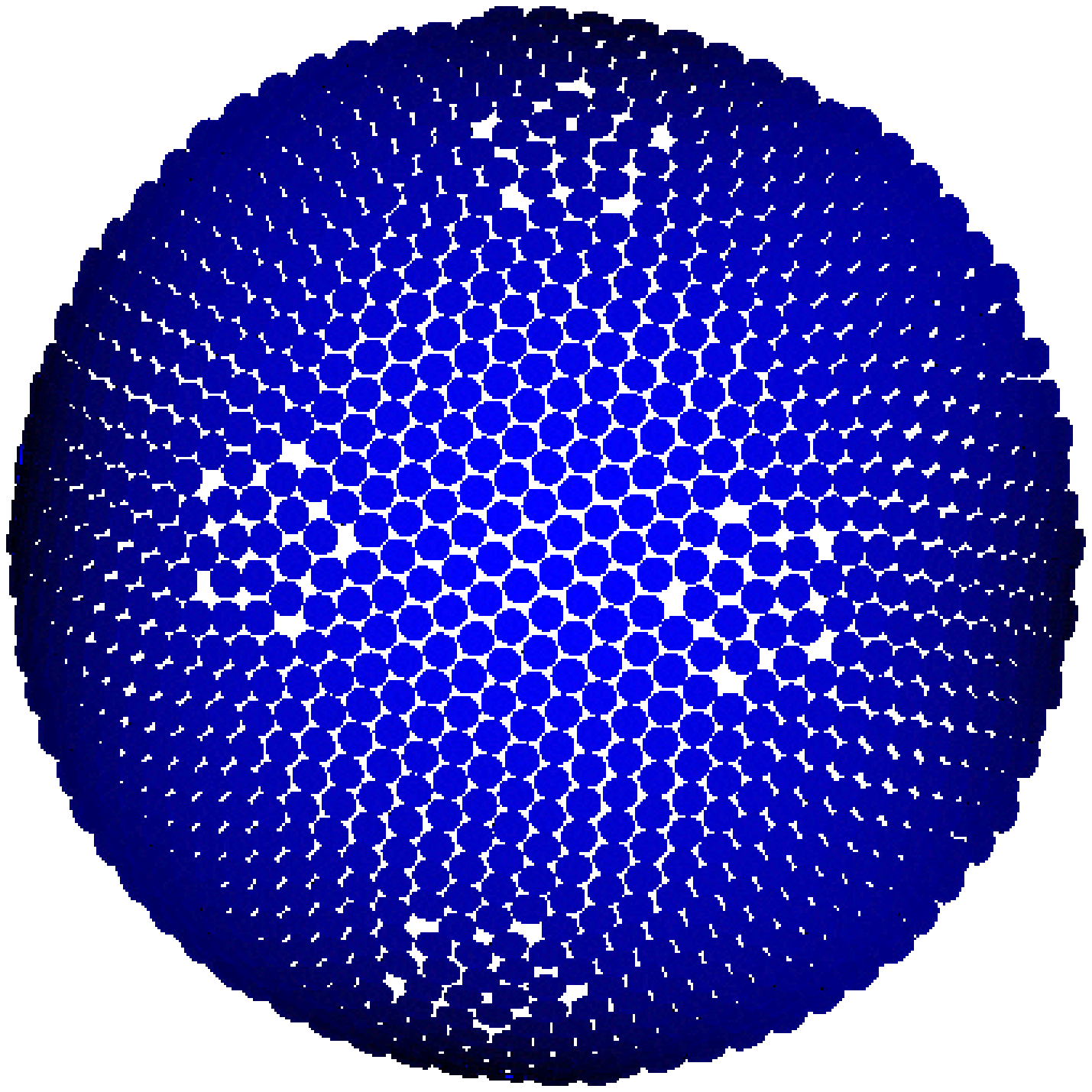}}
\vspace*{-0.05in}
\noindent
Figure 1.
Detector geometry for simulation study. 
\end{figure}
\newpage{}
\clearpage{}
\begin{figure}
\vspace{0.01in}
\centerline{\epsfysize=5.5in
\epsffile{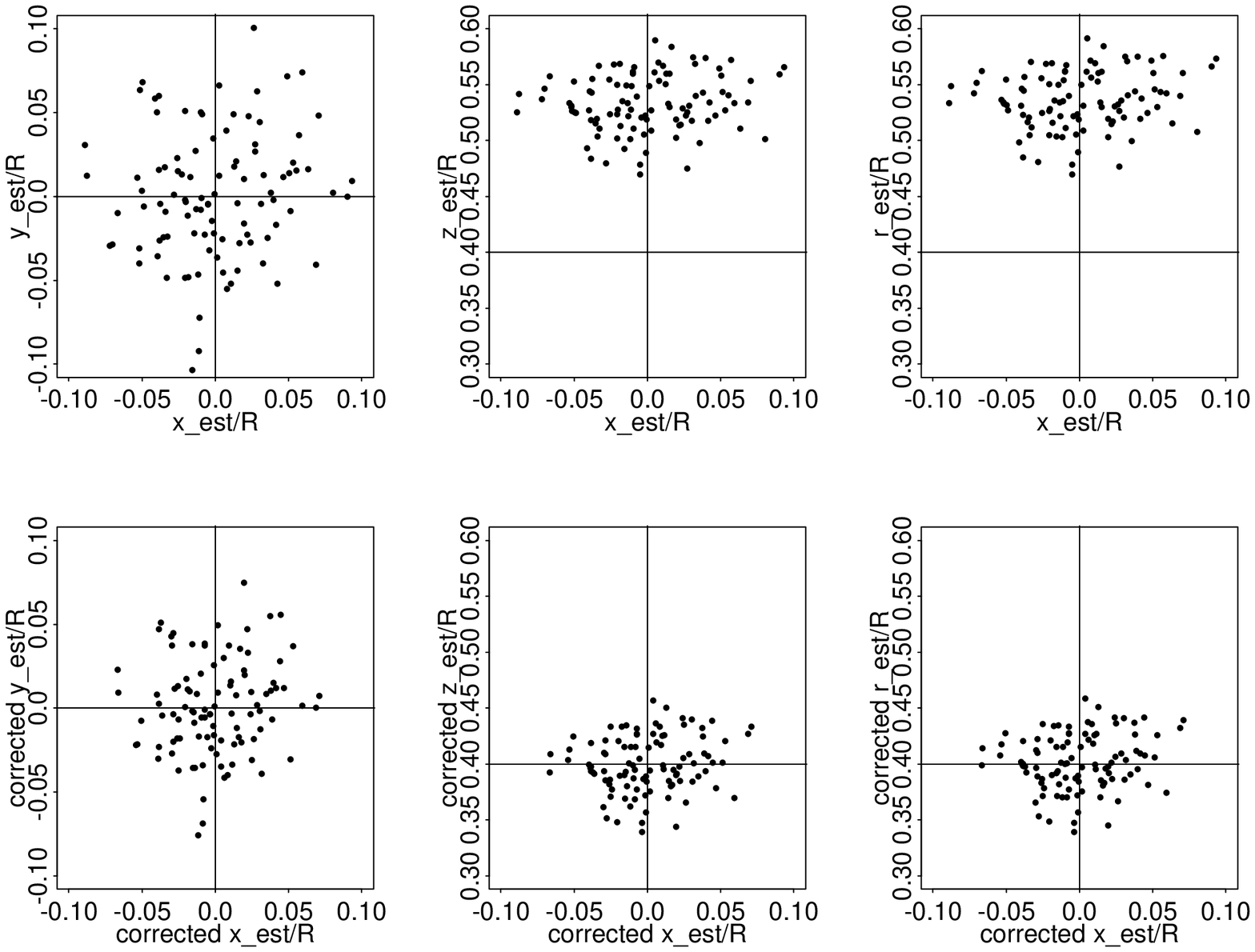}}
\vspace*{-0.05in}
\noindent
Figure 2.
Top row: 100 realizations of the uncorrected x,y,z and radial components of the Maximum Likelihood estimate
of position.
Bottom row: Corrected version of top.
True position is (x,y,z) = (0,0,0.4R).
The scattering length is $\lambda_s = $ 0.1 R.
``No-shift" detection model.
The modified intensity parameter is $\lambda p_e$ = 400.

\end{figure}
\newpage{}
\begin{figure}
\vspace{0.01in}
\centerline{\epsfysize=5.5in
\epsffile{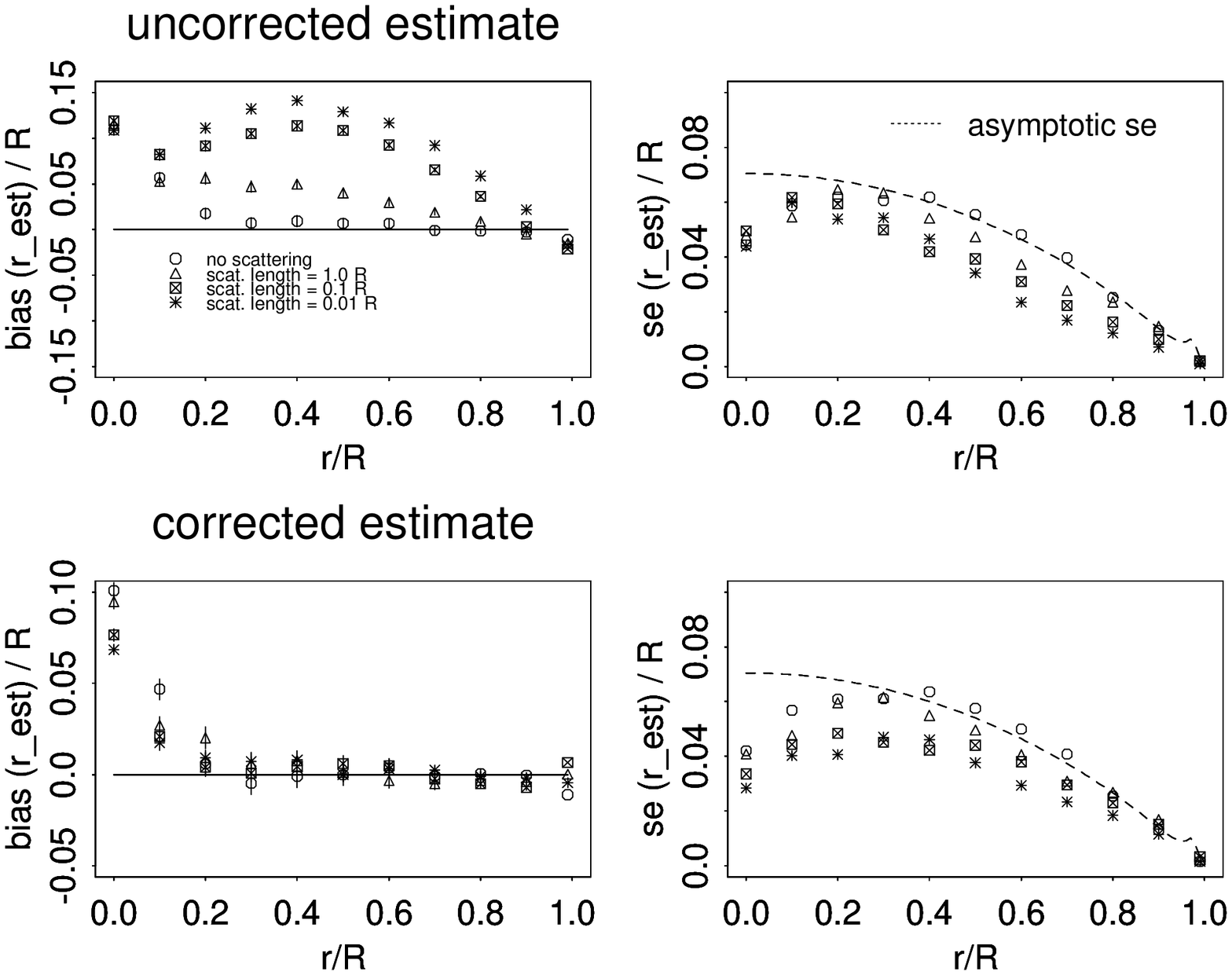}}
\vspace*{-0.05in}
\noindent
Figure 3.
Bias and standard error of corrected and uncorrections radial
estimates of position for data simulated on a grid along the z-axis.
The modified intensity $\lambda p_e $ is 200.
``No-shift" detection model.
\end{figure}
\newpage{}
\begin{figure}
\vspace{0.01in}
\centerline{\epsfysize=5.5in
\epsffile{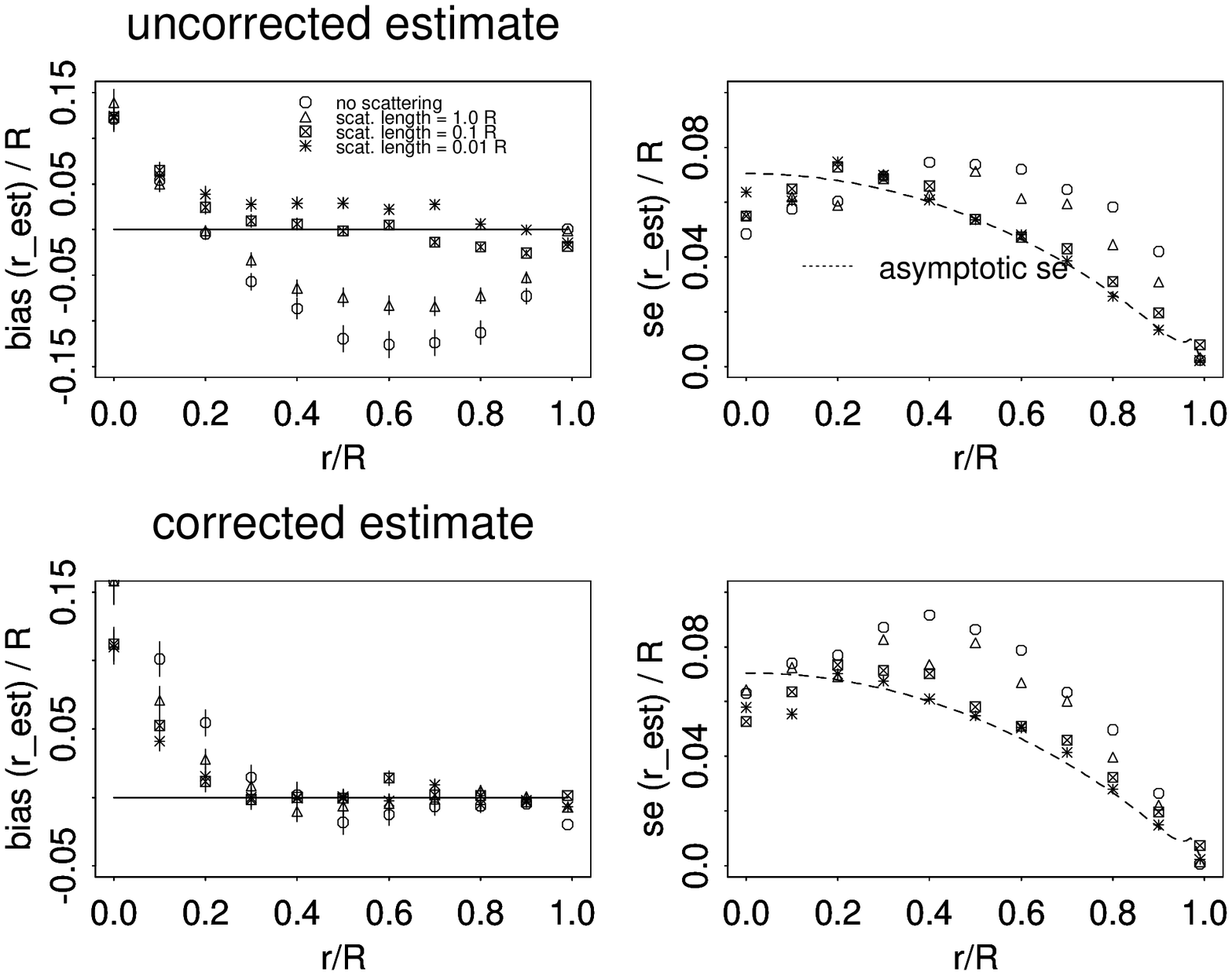}}
\vspace*{-0.05in}
\noindent
Figure 4.
Bias and standard error of corrected and uncorrections radial
estimates of position for data simulated on a grid along the z-axis.
The modified intensity $\lambda p_e $ is 200.
``Shift" detection model.
\end{figure}
\newpage{}

\begin{figure}
\vspace{0.01in}
\centerline{\epsfysize=5.5in
\epsffile{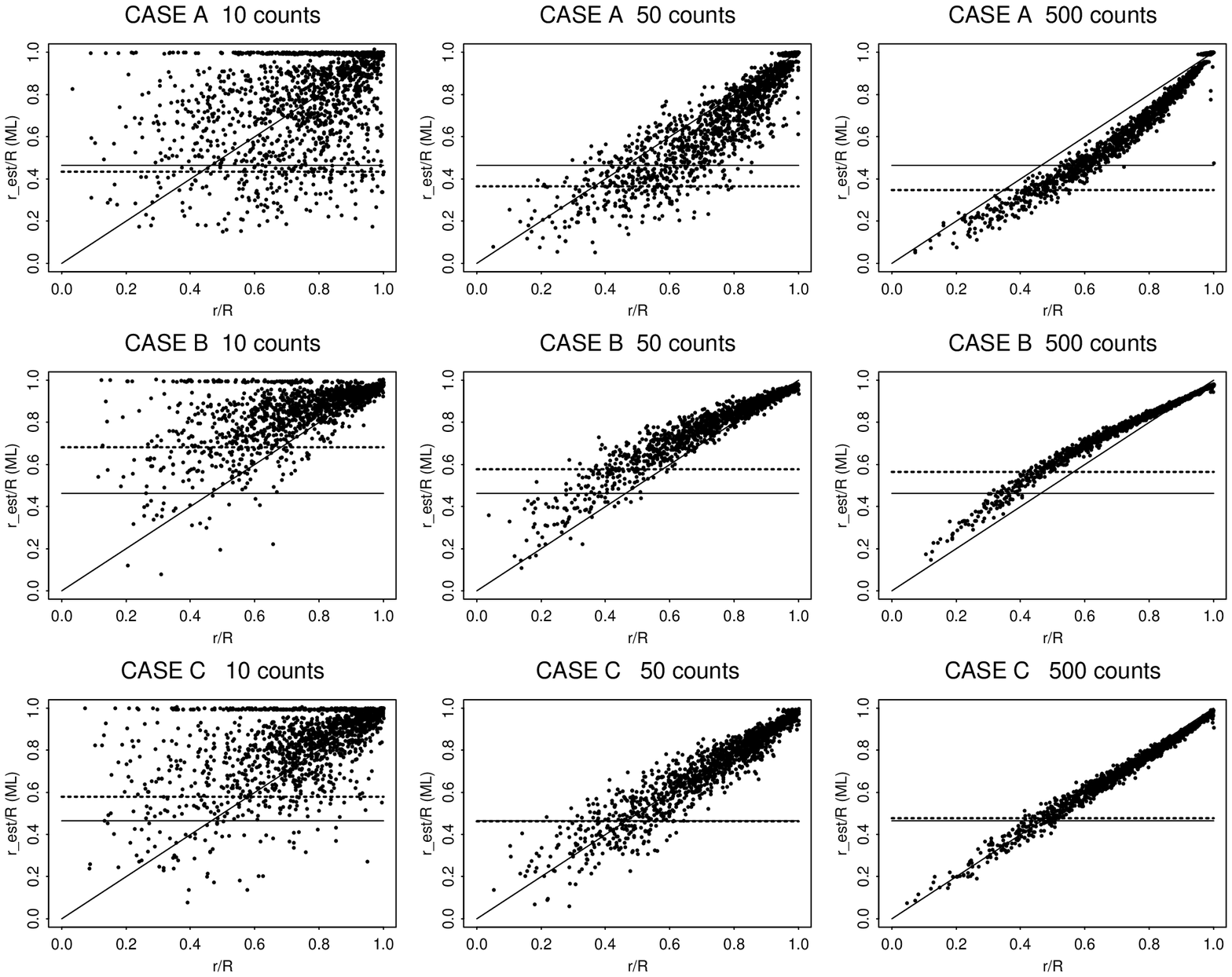}}
\vspace*{-0.05in}
\noindent
Figure 5.
Uncorrected Maximum Likelihood model
estimates of radial position for
events that occur uniformly within the sphere.
CASE A: $\lambda_S = \infty$ and ``shift" detection model
CASE B: $\lambda_S = 0.1 R$ and ``no-shift" detection model.
CASE C: $\lambda_S = 0.1 R$ and ``shift" detection model.
For each plot, we show the 0.1 quantile of
the prediction (dashed line) and 
the 
$r_p$ level. 
\end{figure}
\newpage{}
\begin{figure}
\vspace{0.01in}
\centerline{\epsfysize=5.5in
\epsffile{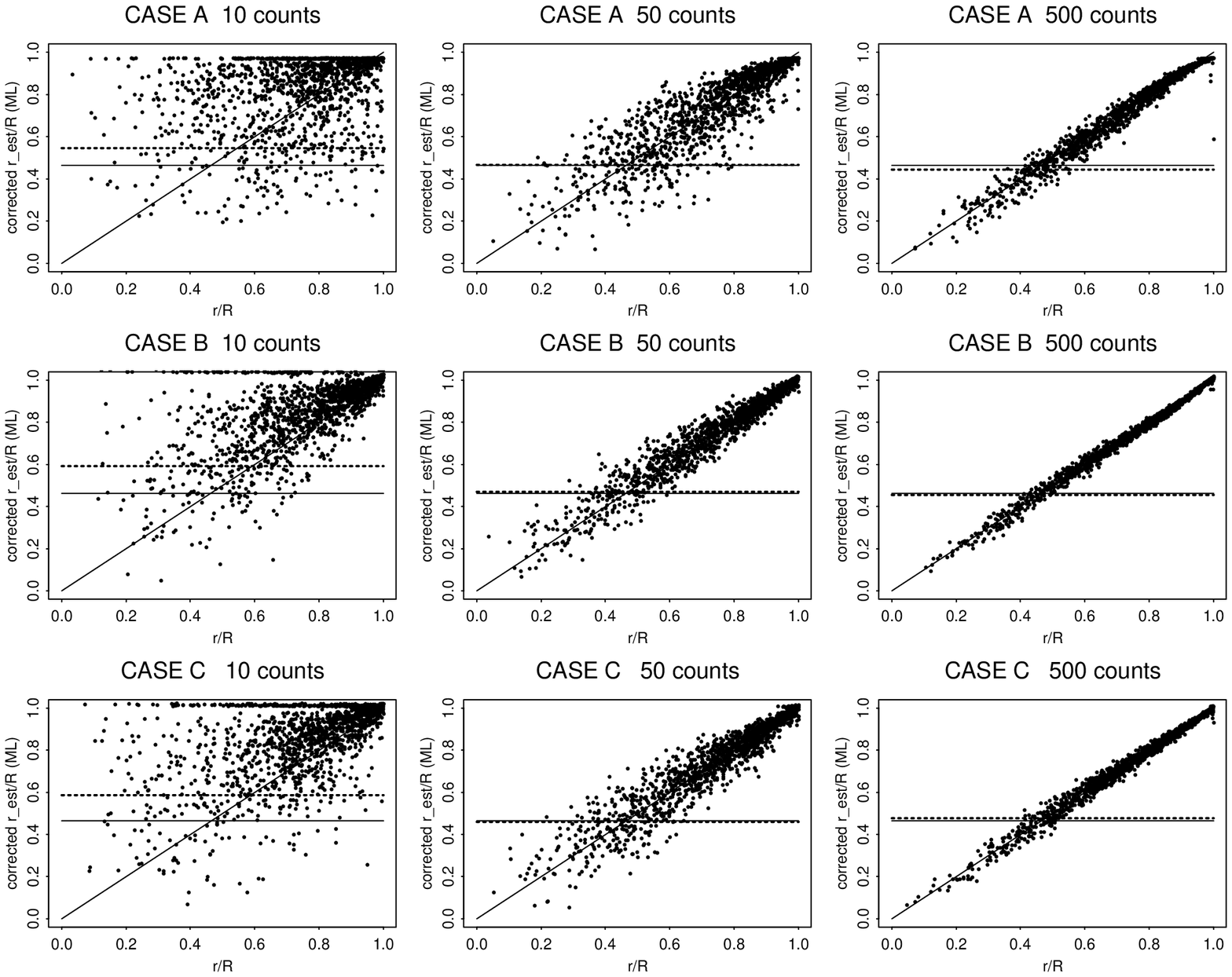}}
\vspace*{-0.05in}
\noindent
Figure 6.
Corrected Maximum Likelihood model
estimates of radial position for
events that occur uniformly within the sphere.
Same data as in Figure 5.
\end{figure}
\newpage{}

\begin{figure}
\vspace{0.01in}
\centerline{\epsfysize=5.5in
\epsffile{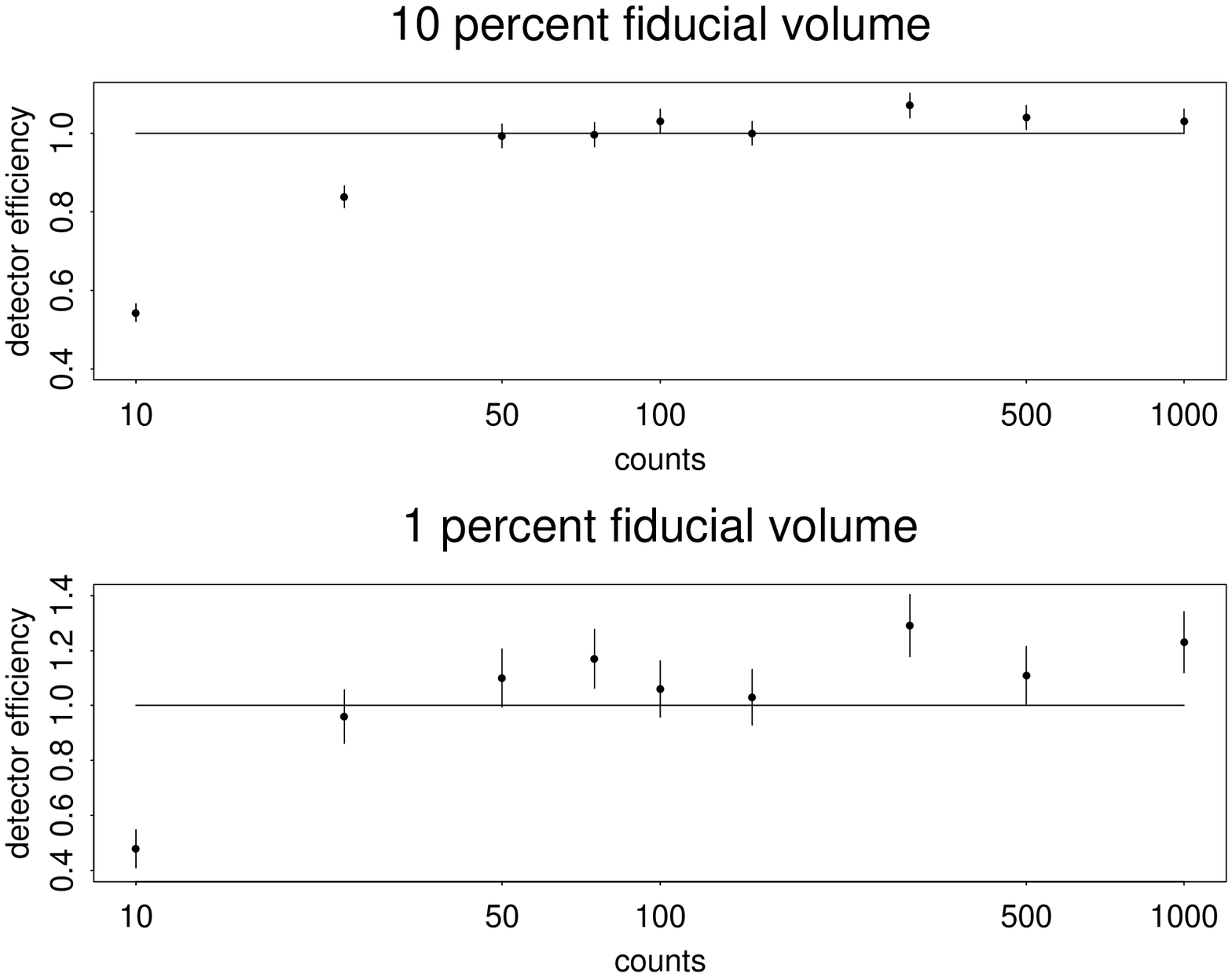}}
\vspace*{-0.05in}
\noindent
Figure 7.
Detector  efficiency for 1 $\%$ and 10
$\%$ fiducial volumes for
$\lambda_s/R = $ 2/9
($\pm$ 1-sigma uncertainty intervals shown).
``Shift" detection model.
\end{figure}

\begin{figure}
\vspace{0.01in}
\centerline{\epsfysize=5.5in
\epsffile{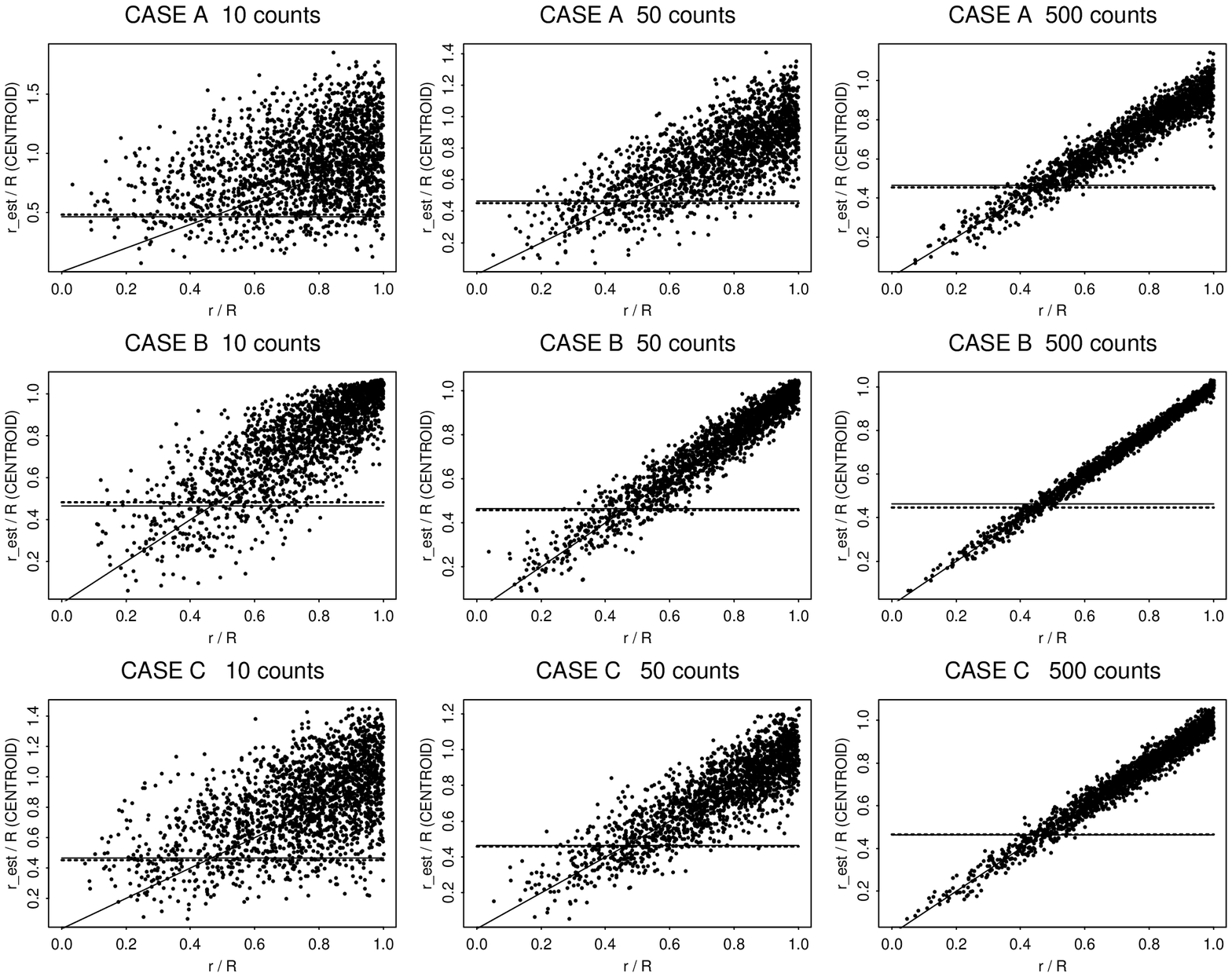}}
\vspace*{-0.05in}
\noindent
Figure 8.
Predicted radial estimate based on
centroid.
Same data as in Figure 5.
\end{figure}
\newpage{}

\begin{figure}
\vspace{0.01in}
\centerline{\epsfysize=5.5in
\epsffile{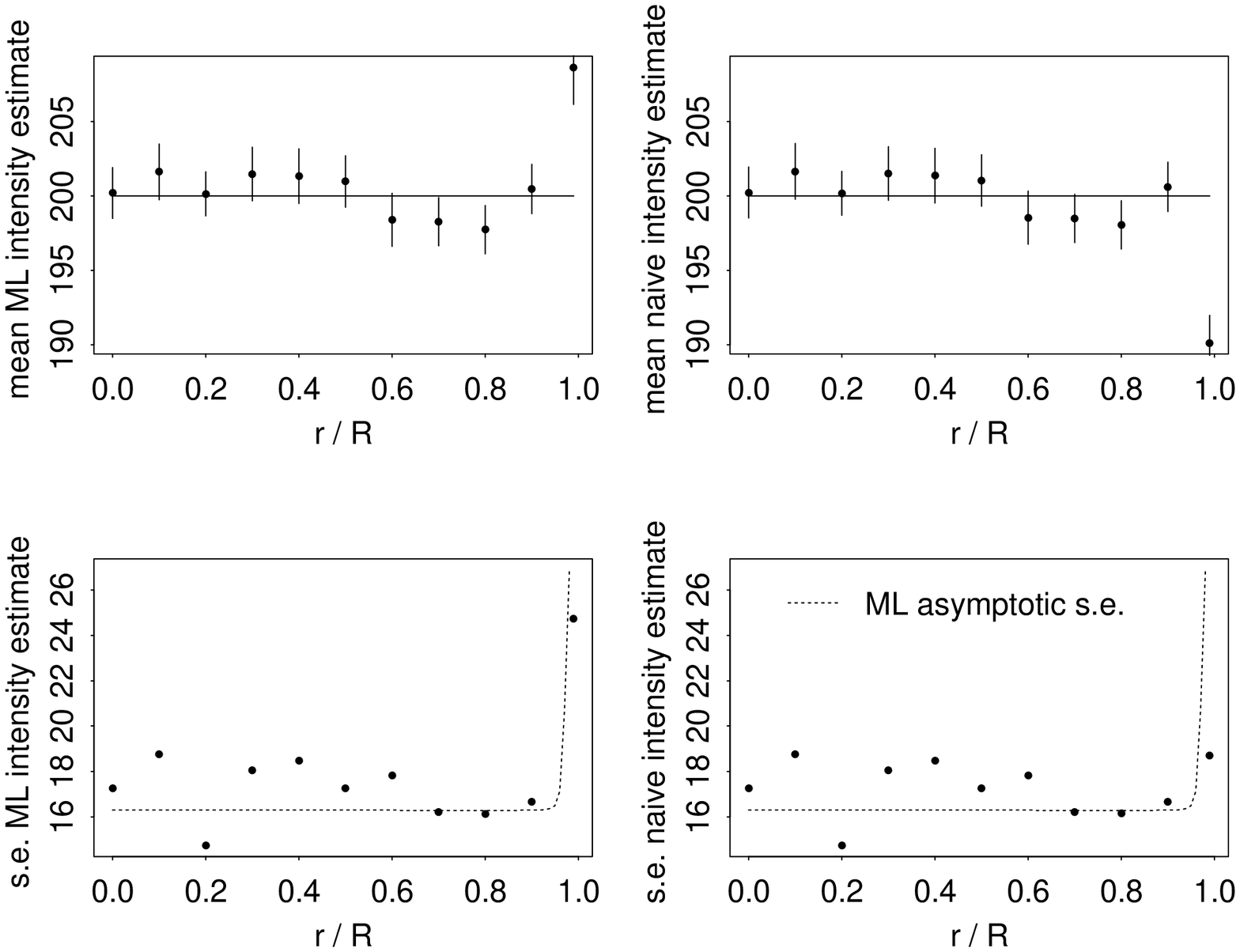}}
\vspace*{-0.05in}
\noindent
Figure 9.
For the ``shift" detection model and $\lambda_s / R = $ 0.1,
we compare the mean and standard error
of two estimates of modified intensity
for the case where event locations
are on the $z$ axis.
The modified intensity is $\lambda p_e=$ 200.
\end{figure}

\begin{figure}
\vspace{0.01in}
\centerline{\epsfysize=5.5in
\epsffile{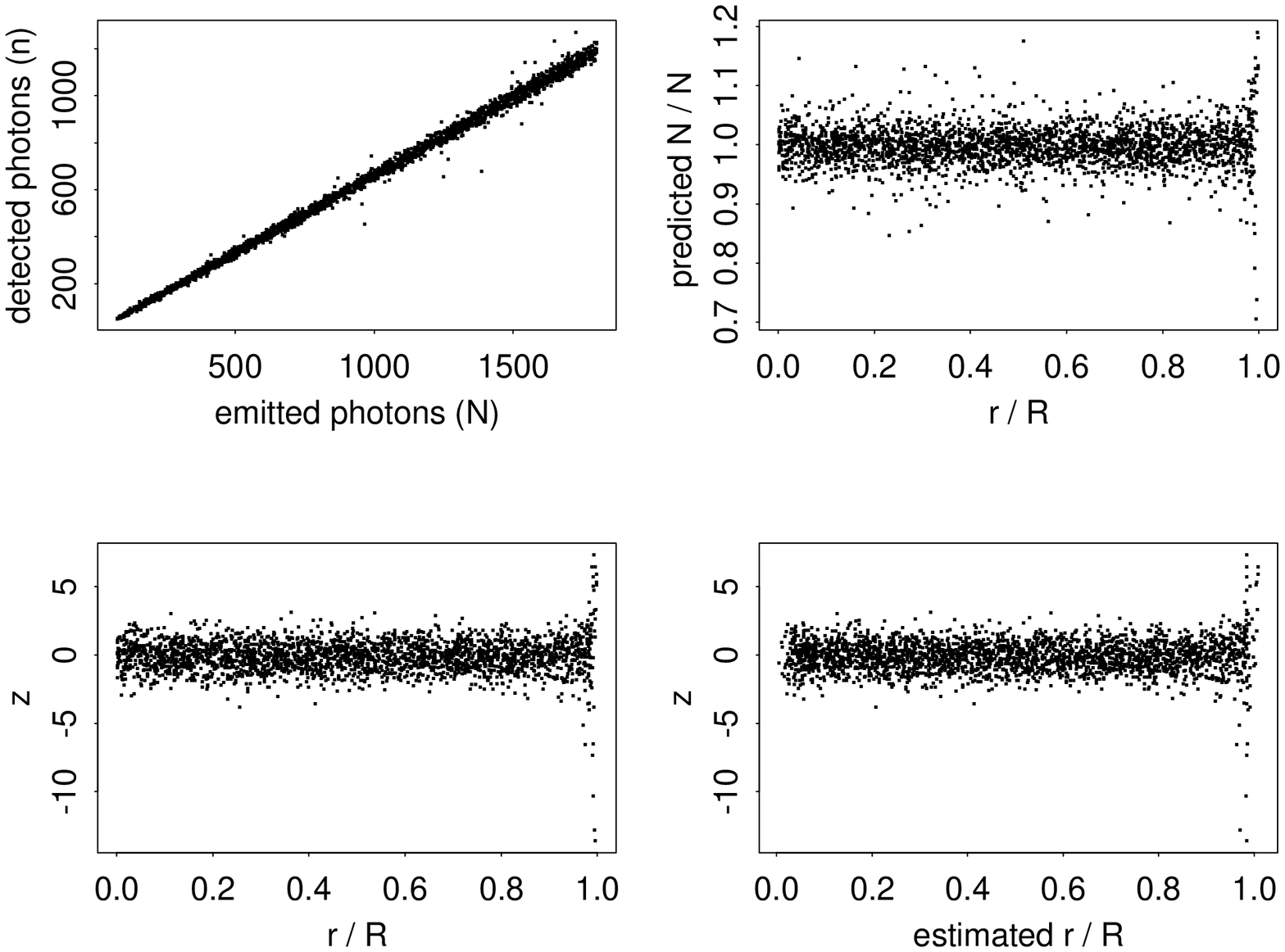}}
\vspace*{-0.05in}
\noindent
Figure 10.
``Shift" detection model and $\lambda_s / R $  = 0.1.
We predict
$N$
based on the number of detected photons $n$
using our ``naive" estimation model.
If our ``naive" model for predicting modified intensity is valid,
$z$ 
should be approximately a Gaussian random variable
with expected value and standard deviation equal to 0 and 1.
\end{figure}
\newpage{}

\end{document}